\documentclass{aa}
\usepackage{epsfig}

\def\ltsima{$\; \buildrel < \over \sim \;$}
\def\simlt{\lower.5ex\hbox{\ltsima}}
\def\gtsima{$\; \buildrel > \over \sim \;$}
\def\simgt{\lower.5ex\hbox{\gtsima}}

\begin{document}
   \thesaurus{03 (13.25.2;11.19.1;11.09.1: NGC~2992)}
  \title{The variability of the Seyfert galaxy NGC~2992: the case for a
revived AGN}

   \author{R. Gilli 
                \inst{1 \star}
   \and    R. Maiolino
		\inst{2}
   \and    A. Marconi
		\inst{2}
   \and    G. Risaliti
                \inst{1}
   \and    M. Dadina
		\inst{3}	
   \and    K.A. Weaver
		\inst{4}
   \and    E.J.M. Colbert
		\inst{4}
}
   \offprints{R. Gilli\\ $^{\star}$ Present address: Astrophysikalisches 
	Institut Potsdam, An der Sternwarte 16, D--14482 Potsdam, Germany}

  \institute {Dipartimento di Astronomia e Scienza dello Spazio, Universit\`a 
di Firenze, Largo E. Fermi 5, I--50125 Firenze, Italy
(gilli, risaliti@arcetri.astro.it)
     \and   Osservatorio Astrofisico di Arcetri, Largo E. Fermi 5,
I--50125 Firenze, Italy (maiolino, marconi@arcetri.astro.it)
     \and BeppoSAX SDC, ASI, Via Corcolle 19, I--00131 Roma, Italy
(dadina@gavi.sdc.asi.it)
     \and Laboratory for High Energy Astrophysics, Code 662, NASA/GSFC,
Greenbelt, MD 20771, USA (kweaver, colbert@gsfc.nasa.gov) 
}

   \date{Received / Accepted }

\titlerunning{Variability of NGC~2992}
\authorrunning{R. Gilli et al.}

   \maketitle

   \begin{abstract}

We report the transition to an active state of the nucleus
in the Seyfert 1.9 galaxy NGC~2992, discovered by means of new
hard X--ray data. While the 2--10 keV flux declined by a factor
of $\sim$ 20 from 1978 to 1994, two recent BeppoSAX observations in 1997
and in 1998 caught the nuclear emission raising back to the same level of 
activity observed in 1978. 

In both BeppoSAX observations the X--ray spectrum of the source is 
well represented by a power law  with spectral index $\Gamma\simeq 1.7$, 
absorbed by a column density of $N_{\rm H}\simeq 10^{22}$ cm$^{-2}$  
and characterized by a prominent iron K$\alpha$ line. 
While in the second BeppoSAX data set the line properties appear to be 
consistent with those expected from accretion disc models,
in the first BeppoSAX data set the iron feature is rather peculiar. 
The broadening is not significant and 
the line energy is $E_{\rm K\alpha}=6.62\pm0.07$ keV,
indicating emission from highly ionized iron. The line has too high 
equivalent width ($\sim 700$ eV) to be produced by a hot scattering medium.
By comparing these data with data previously in the literature, 
we interpret the spectral and flux changes in terms of different phases 
of rebuilding an accretion disc. The timescale for the
disc rebuilding is estimated to range between 1 and 5 years. 


The X--ray data are complemented with optical and near-infrared followup
spectra taken 1.5 months after the discovery of the X--ray burst.
The spectra are characterized by prominent broad emission lines.
There is also
evidence for hot dust emission in the H and K bands that, however, is
probably still in the process of increasing.

  \end{abstract}


%

\section{Introduction}

The Seyfert 1.9 galaxy NGC~2992 is a nearby (z=0.0077), edge on
($i=70^{\circ}$) system interacting with NGC~2993.
The detection of a broad
component of H$\alpha$ with no H$\beta$ counterpart (Ward
et al. 1980; Shuder 1980) implied an intermediate Seyfert type
classification, suggesting the existence of an obscured
broad line region (BLR). This was confirmed later by the infrared
detection  
of a broad Pa$\alpha$ line (Rix et al. 1990; Goodrich, Veilleux \& Hill
1994). The optical image of NGC~2992 (Ward et al. 1980) shows a prominent
dust lane extending along the major axis of the galaxy and crossing the
nucleus.
The optical extinction, measured from the Balmer decrement of the
narrow H$\alpha$ and H$\beta$ components, is $A_V=4.2\pm0.4$ (Ward et al.
1980). 
Finally, evidence for coronal lines of [FeVII] and [FeX]
have been
found in the optical spectrum by Shuder (1980) and Winkler (1992). 

The X--ray flux and spectrum were observed to vary during 16
years of observations performed with different satellites. 
As noted by Weaver et al. (1996; hereafter We96), 
since the time of HEAO--1 observations in 1978 (Mushotzky 1982), the 2--10
keV flux of NGC~2992 decreased by a factor of $\sim 20$ until the ASCA 
observations in 1994,
when it reached the lowest value. A similar decrease was
also observed in the soft X--rays (We96). Before ASCA,
the spectral shape exhibited some variations but on
average it could be
described as an absorbed ($N_{\rm H}\sim 10^{22}$ cm$^{-2}$) power law
with photon index $\Gamma \sim 1.7$ and an iron K$\alpha$ line. The ASCA
spectrum was however much different, with a very flat slope ($\Gamma \sim
1.2$). These spectral variations have been
interpreted by We96 as the evidence for radiation reprocessed by an 
optically and geometrically thick gaseous medium, e.g. the molecular torus  
invoked by unified models (Antonucci \& Miller 1985).
We96 measured a time lag between the Fe line and the continuum of 10$\pm$4 
years, from which
they estimate a distance of the torus from
the central engine ($\sim 3.2$ pc, that however is still model-dependent). 
Flux variability in the infrared was found 
by Glass (1997), who in particular remarked a fading of
the source from 1978 to 1996 apart from a strong outburst in 1988. 
All of this evidence suggests that before BeppoSAX observed NGC~2992 in 
1997 and 1998, the AGN in the center was turning off. 

In this paper we present an analysis of new BeppoSAX
observations of NGC~2992 which show that the AGN is back to its previous  
high level of 
activity observed in 1978. The X--ray data are also complemented with
optical and near--IR spectra obtained in a followup observation. The
results of
our analysis are compared with previous data in the literature.

We assume $H_0=50$ km s$^{-1}$ Mpc$^{-1}$ yielding a distance of 46 Mpc
for NGC~2992. 

\section{X--ray data}\label{xdata}

\subsection{New observations and data reduction}

The four co--aligned Narrow Field Instruments (NFI) on board
the Italian--Dutch satellite BeppoSAX (Boella et al. 1997a) span the 
broad X--ray energy range from 0.1 to 200 keV. The NFI include
the Low Energy Concentrator Spectrometer (LECS; Parmar et al. 1997) and
the Medium Energy Concentrator Spectrometer (MECS; Boella et al. 1997b),
which both consist of a mirror unit plus a gas scintillation proportional
counter and have imaging capabilities. The LECS instrument covers the
energy range 0.1--10 keV with a spatial resolution of about $\sim 1$
arcmin and a spectral resolution of 8\% at 6 keV.  The MECS instrument
operates in the 1.8--10 keV band with the same spatial and spectral
resolution of the LECS, but is two times more sensitive in the overlapping 
spectral range. The other NFI are
a High Pressure Gas Proportional Counter (HPGSPC; Manzo et al. 1997) and a
Phoswich Detector System (PDS; Frontera et al. 1997), which are
direct--view detectors with rocking collimators. The HPGSPC (4--120 keV)
has better energy resolution but is less sensitive than the PDS (13--200
keV). In this paper we will only take advantage of the LECS, MECS
and
PDS data. Errors, unless otherwise stated, will be given at the 90\%
confidence level for one interesting parameter ($\Delta\chi^2=2.71$).

NGC~2992 was observed by BeppoSAX in 1997 (from December 1 to December
3) and in 1998 (from November 25 to November 27). The 1997 data
and the 1998 data will be referred to as SAX1 and
SAX2, respectively. For the SAX1 observation we will not take into account
the LECS data due to poor statistics. Table~\ref{obs_log} shows exposure
times and net count rates (i.e. background subtracted) for both BeppoSAX
observations. The MECS spectra for SAX1 and SAX2, and the LECS spectrum
for SAX2 have been extracted with an aperture radius of 4 arcmin. The
background subtraction files have been extracted with the same apertures
from blank sky fields. The spectral data have been binned in order to have
at least 20 counts per channel to validate the use of the $\chi^2$
statistics.

A possible source of contamination in the X--ray spectrum detected by
BeppoSAX could be NGC~2993, which lies at about 2.9 arcmin from
its companion NGC~2992.
NGC~2993 has an optical magnitude comparable
to that of NGC~2992 ($\Delta m=0.03$) and also shows optical narrow
emission
lines and radio emission (Ward et al. 1980). On the basis of the optical
lines, NGC~2993 has been classified as a LINER (Durret 1990). Due to
the uncertain nature of LINERs, which could host both thermal and 
non-thermal activity, NGC~2993 could contribute to the flux measured by 
BeppoSAX. However, since at 1415 MHz NGC~2993 is four times less luminous
than NGC~2992 (Ward et al. 1980), and radio emission should 
not suffer from absorption, NGC~2993 should be intrinsically less luminous
than NGC~2992 and its contribution to the observed
X--ray spectrum should be small in
every X--ray energy band from LECS to PDS.
Also, the ROSAT PSPC data show that
in the soft X--rays the flux of NGC~2993 is less than 5\% of the
NGC~2992 flux (We96), 
and an analysis of the ASCA SIS image of the NGC~2992 field has
revealed that the emission of NGC~2993 is negligible also in the 2--10 keV
band. 
Therefore, we consider the X--ray photons detected by BeppoSAX
as entirely produced by NGC~2992.

\begin{table}
\caption{BeppoSAX observations of NGC~2992.}
\label{obs_log}
\begin{tabular}{lcc}
\hline \hline   
NFI$^a$& Exposure time$^b$& Net count rate$^c$\\
\hline 
\multicolumn{3}{l}{\it SAX1(1997)}\\ 
\hline
MECS& 72119& $0.082\pm0.001$\\
PDS& 33820& $0.226\pm0.048$\\  \hline 
\multicolumn{3}{l}{\it SAX2(1998)}\\
\hline  
LECS& 22284& $0.411\pm0.004$\\
MECS& 59245& $0.893\pm0.004$\\
PDS& 27084& $1.405\pm0.005$\cr\hline
\end{tabular}

Errors on the counts are at $1\sigma$ level.\\ \\
$^a$Narrow Field Instrument.\\
$^b$In seconds.\\
$^c$Background subtracted.\\
\end{table}

\subsection{Timing analysis}

\begin{figure}
\epsfig{file=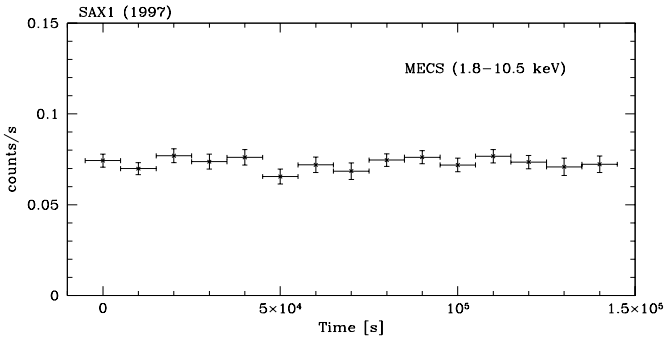, width=9.cm, height=5cm,
angle=0}
\caption{Light curve for the MECS (1.8--10.5 keV) instrument during the
SAX1 observation. The binning
time is 10 ks; the light curve is not background subtracted.}
\label{sax1timing}
\end{figure}

The flux of NGC~2992 was constant throughout the SAX1
observation in 1997 (Fig.~\ref{sax1timing}). When a fit with a constant
is performed to the SAX1 light curve, the result is $\chi^2/dof=9.4/14$ for
a binning time of 10 ks.  
On the contrary the source exhibited short term variability in 1998 during
the SAX2 observation, when it was in the high state: the source
varied by a factor of $\sim 1.4$ on timescales of
$7\times10^4$ s. A constant count rate is ruled out at $>99\%$ confidence
level by the $\chi^2$ test both for the LECS and MECS data. 

The variability amplitude of NGC~2992 during the SAX2 observation was
estimated by means of the normalized ``excess variance'' $\sigma^2_{rms}$ 
(see Nandra et al. 1997a for the definition of this quantity). 
In order to allow a comparison with the results of Nandra et al. (1997a)
we have calculated $\sigma^2_{rms}$ with a binning time of 128 s,
which also ensures Gaussian statistics. For the 2--10 keV lightcurve
$\sigma^2_{rms}$ turned out to be $9.7(\pm0.3)\times10^{-3}$. 
This value is typical of Seyfert 1s with the same intrinsic 2--10 keV
luminosity of NGC~2992 (Nandra et al. 1997a).
We note that
the value of $\sigma^2_{rms}$ depends on the duration of the
observation and that the exposures times for the Seyfert 1s sample of
Nandra et al. (1997a) are all shorter than $\sim 60$ ks, the MECS
exposure time in SAX2. Therefore, in order to be consistent with the
average exposure of the Nandra et al. (1997a) sample, we have considered
different time intervals of $\sim 30$ ks (first and second half of the
observation and a central interval). We obtain a mean $\sigma^2_{rms}$ of 
$6.7(\pm0.2)\times10^{-3}$, which still agrees with that observed in 
Seyfert 1s of comparable luminosity.

\begin{figure}   
\epsfig{file=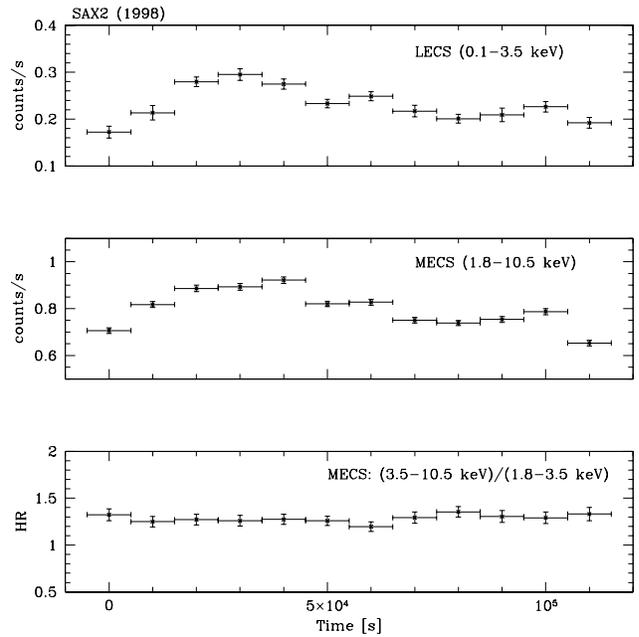, width=9.cm, height=9.cm,
angle=0}
\caption{Light curves for the LECS (0.1--3.5 keV) and MECS (1.8--10.5 keV)
instruments and Hardness Ratio between MECS data during the SAX2
observation.
The binning time is 10 ks; the light curves are not background
subtracted.}
\label{sax2timing}
\end{figure}

In order to
check for spectral variations associated with the SAX2 short 
term variability, we have analyzed the ratio between the hard and soft
X--ray photons (Hardness Ratio, HR) in different bands. 
When the HR is calculated comparing LECS to MECS data, one could not
exclude some degree of spectral variability. However, HR variations are
not detected when the HR is calculated only from MECS data.
As an example (see Fig.~\ref{sax2timing}), the HR between the MECS data in
the 3.5--10.5
keV and 1.8--3.5 keV is constant at a high confidence level
($\chi^2/dof=6.1/11$).    
We note that the cross calibration between the LECS and MECS data
is somewhat uncertain (see the next Section), and it might exhibit time
variations. Therefore, we will consider the dispersion in the HR
between the LECS and MECS data as an instrumental effect and extract 
only a time averaged spectrum. 

\subsection{Spectral analysis}

{\sl SAX1 (1997)}\\

\begin{figure}
\epsfig{file=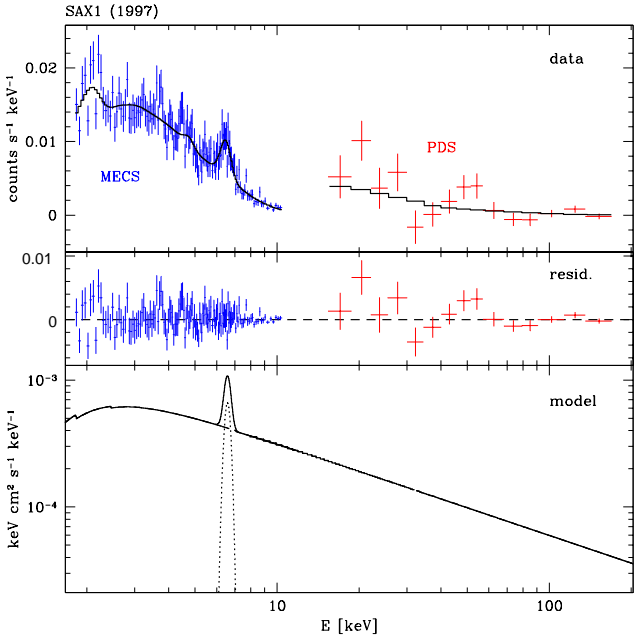, width=9.cm, height=9.cm,
angle=0}
\caption{Data and folded model (top), residuals (middle) and unfolded
model (bottom) for the SAX1 observation.}
\label{sax1_spectrum}
\end{figure}

The spectral analysis was performed by using the XSPEC 10.0 package.
The MECS data in the 1.8--10.5 keV band and the PDS data in the 14--200
keV band 
were fitted simultaneously. A relative normalization factor was 
introduced between the PDS and MECS spectrum since the BeppoSAX instruments 
show some mismatches in the absolute flux calibration (the MECS is
the best calibrated instrument). The energy ranges covered by the
MECS and PDS do not overlap and so we cannot determine a
cross calibration constant from the data. Instead, for the fitting
procedure, we assumed a fiducial PDS to MECS normalization factor of
0.8 (released by the BeppoSAX Science Data Center).
When fitting with an absorbed power law, strong residuals emerge at
6--7 keV suggesting a prominent iron line in the NGC~2992 spectrum.
With the addition of a Gaussian iron line to the absorbed
power law model we obtain a good fit to the data with $\chi^2/dof=149/144$.
The fit is shown in Fig.~\ref{sax1_spectrum} and spectral parameters
are quoted in Table~\ref{results} together with the hard X--ray
flux and luminosity derived by the best fitting model. 
Spectral parameters are quoted in the rest frame. 

\begin{figure}
\epsfig{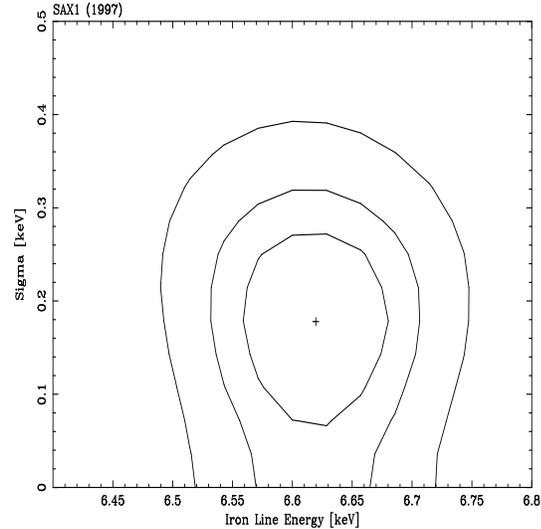}   
\caption{Confidence contours of the iron line energy versus the line
width for the SAX1 observation. The contour levels are 68\%, 90\% and
99\% for two interesting parameters.} \label{sax1_contour}
\end{figure}

The power law spectral index ($\Gamma=1.72\pm0.12$) is
consistent with the
typical value 1.7--1.9 for Seyfert galaxies (Nandra \& Pounds 1994;
Nandra et al. 1997b). 
The line broadening ($\sigma_{\rm K\alpha}=0.18^{+0.10}_{-0.14}$ keV) 
is not highly significant (it is detected at $\stackrel{>}{_{\sim}}90\%$ 
confidence level, 
$\Delta\chi^2=3.21$ with respect to a narrow fit with 
$\sigma_{\rm K\alpha}=0$ keV).
On the other hand, the energy of the Fe line,
$E_{\rm K\alpha}=6.62\pm0.07$ keV, is significantly higher ($>99.9\%$ 
confidence level, $\Delta\chi^2>21$) than the value of 6.4 keV commonly 
observed in Seyfert
galaxies (e.g. Nandra et al. 1997b; Turner et al. 1997) and 
expected from neutral or weakly ionized
iron (below FeXVI).  We note, however, that both
the MECS2 and the MECS3 units (the MECS1 failed in
May 1997) might be affected by an instrumental energy shift towards
the blue of 0.05 and 0.08 keV at 6.6 keV respectively 
(SAX Cookbook: 
$\rm http://www.sdc.asi.it/software/cookbook/cross\underline{~}cal.html$).
We have
verified the significance of our results by calculating the confidence
contours for the line energy versus the line width (shown in
Fig.~\ref{sax1_contour}) and versus the photon index (not shown). 
Even applying a red shift of 0.05 or 0.08 keV to correct for the instrumental
shift we find that $E_{\rm K\alpha}>6.4$ keV at $>99\%$ confidence level.
Therefore, evidence for iron emission from a highly ionized
medium is provided by the data.
The equivalent width of the line is very high,
$EW_{\rm K\alpha}\sim 700$ eV and cannot be easily explained with current
emission line models for a direct view of the nucleus (see
Section~\ref{discussion}).

We note that the PDS data are very noisy and do not allow to constraint
significantly any reflection component in addition to the power law
continuum.\\

\begin{table*}
\caption{Spectral fits to the BeppoSAX data}
\label{results}
\begin{tabular}{ccccccccc}
\hline \hline
Model& $\Gamma$& $N_{\rm H}$& $E_{\rm K\alpha}$& $\sigma_{\rm
K\alpha}$&
$EW_{\rm K\alpha}$& $f_{\rm 2-10\,keV}$& $L_{\rm 2-10\,keV}$&
$\chi^2/dof$\\
& & $[10^{22}{\rm cm}^{-2}]$& [keV]& [keV]& [eV]& $[10^{-11}{\rm
erg~s}^{-1}{\rm cm}^{-2}]$& $[10^{42}\rm{erg~s}^{-1}]$\\ 
\hline
\multicolumn{9}{l}{\it SAX1(1997)}\\
\hline
apl+Gl~~~~& $1.72^{+0.13}_{-0.12}$& $1.4^{+0.5}_{-0.4}$& 
$6.62^{+0.07}_{-0.07}$& $0.18^{+0.10}_{-0.14}$& $701^{+185}_{-163}$&
0.63& 1.8& 149/144\\
\hline
\multicolumn{9}{l}{\it SAX2(1998)}\\ \hline
apl+Gl~~~~& $1.70\pm0.02$& $0.90\pm0.03$&
$6.59^{+0.13}_{-0.15}$& $0.32^{+0.18}_{-0.06}$&  $147\pm37$& 7.4& 20&
390/371\cr
\hline
\end{tabular}

apl+Gl: absorbed power law plus Gaussian line. The luminosities are
corrected for the absorption.
\end{table*}

{\sl SAX2 (1998)}\\

The spectral data from the LECS, MECS and PDS were fitted
simultaneously in the 0.1--3.5 keV, 1.8--10.5 keV and 14--200 keV bands,
respectively. As for the SAX1 observation, the PDS to MECS normalization
was fixed to 0.8. The LECS to MECS normalization was determined by
fitting simultaneously the LECS and MECS data in the overlapping band
1.8--3.5 keV; the obtained value was subsequently assumed for the
spectral analysis. The procedure provides a model--independent  
LECS to MECS normalization of $0.64\pm0.02$.
The typical values observed so far range from 0.65 (Malaguti et al.
1999) to 0.76 (Haardt et al. 1998; Guainazzi et al. 1999).\\ 

{\sl Fit with a power law and a Gaussian line}\\

\begin{figure}
\epsfig{file=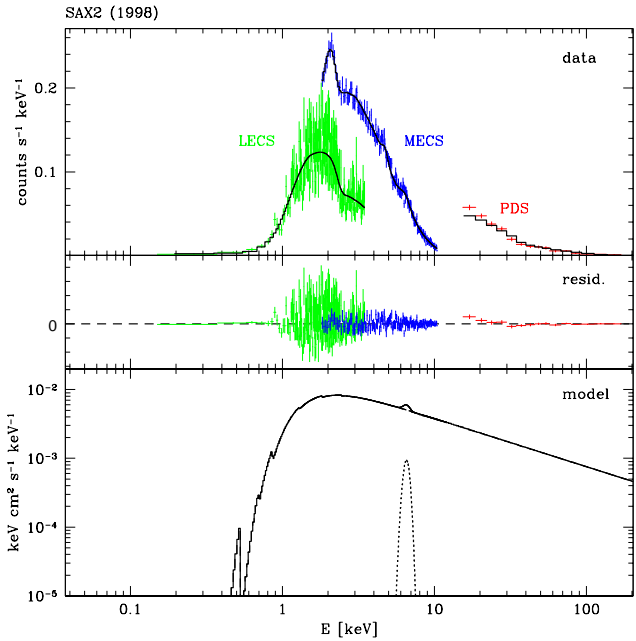, width=9.cm, height=9.cm,
angle=0}
\caption{Data and folded model (top), residuals (middle) and unfolded
model (bottom) for the SAX2 observation.}
\label{sax2_spectrum}
\end{figure}

A simple absorbed power law model is unacceptable at $>97\%$ 
confidence level ($\chi^2/dof=431/374$) mostly because of the strong 
residuals at 6--7 keV due to 
the iron emission. The addition of a broad Gaussian
iron line improves significantly the fit ($\Delta\chi^2\simeq 41$).
The spectral parameters and fit are shown in
Table~\ref{results} and Fig.~\ref{sax2_spectrum}, respectively.
Some residuals at 15--20 keV could suggest
that a reflection component is present. 
However, including a reflection continuum, obtained 
by means of the {\tt pexrav} model in XSPEC (Magdziarz \& Zdziarski 1995),
is not strongly required by the data ($\Delta\chi^2\sim1$ with the
addition of one $dof$). The relative normalization $R$ between the
reflected and the direct continuum turns out to be low, $R=0.12$
($R\sim1$ for a reflecting material covering $2\pi$ of the source) and the
steepening of the spectral index, which should be observed when adding a 
reflection continuum, is only $\sim 1\%$.\footnote {We verified that
the
relative normalization of the reflection component is low ($R=0.17$)
also in the case of a ionized reflector, obtained
by means of the {\tt pexriv} model in XSPEC (Magdziarz \& Zdziarski
1995).}

\begin{figure}
\epsfig{file=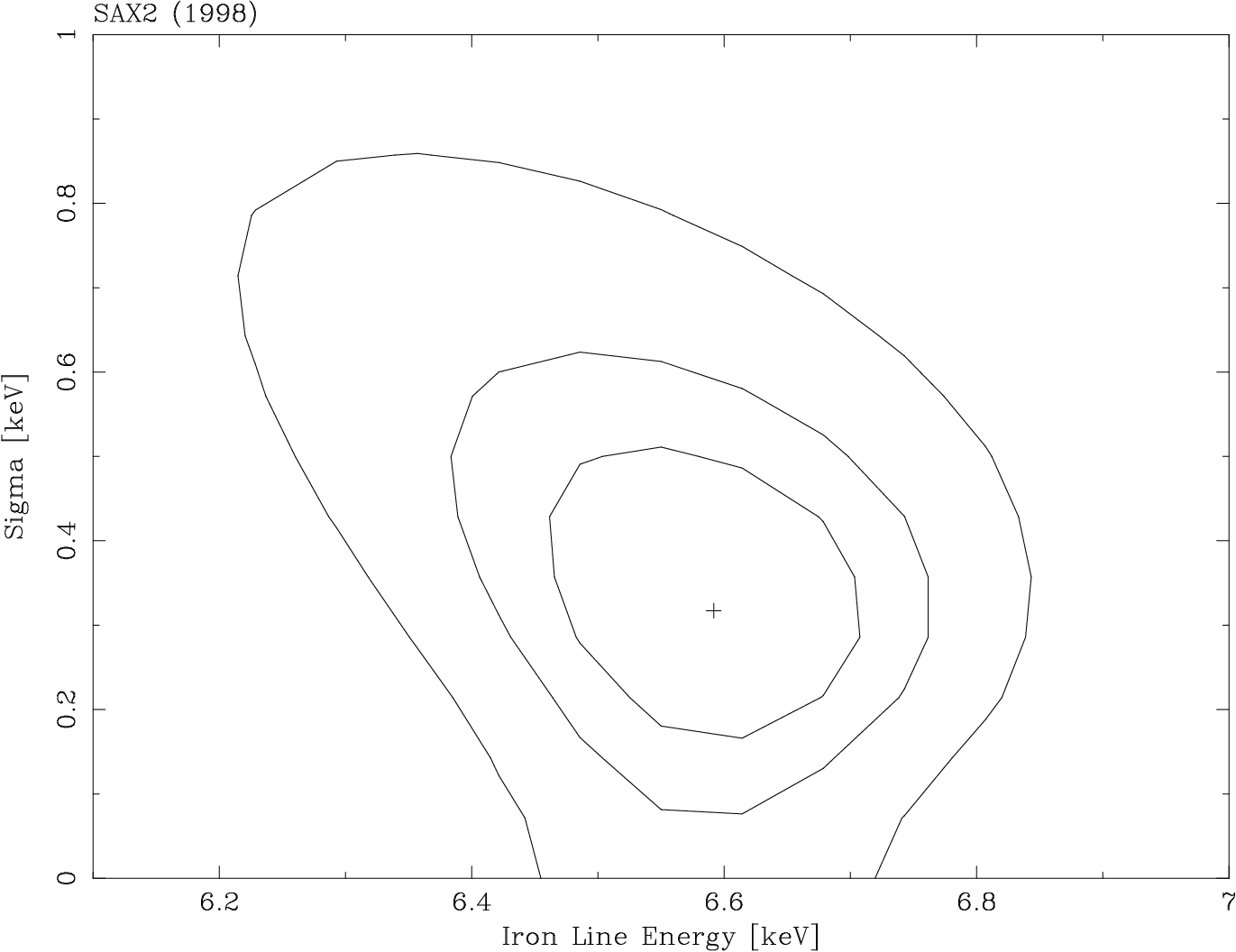, width=7.cm, height=7.cm,
angle=270}
\caption{Confidence contours of the iron line energy versus the line     
width for the SAX2 observation. The contour levels are 68\%, 90\% and
99\% for two interesting parameters.} \label{sax2_contour}
\end{figure}

\begin{table*}
\caption{The iron line complex in the BeppoSAX data.}
\label{irontab}
\begin{tabular}{ccccccc}
\hline \hline
Model& $E_{\rm K\alpha}$& $\sigma_{\rm K\alpha}$&
$EW_{\rm K\alpha}$& $A_{\rm K\alpha}$& $\theta$& $\chi^2/dof$\\
& [keV]& [keV]& [eV]& $[10^{-5}{\rm ph~s}^{-1}{\rm cm}^{-2}]$&
[deg]& \\
\hline
\multicolumn{6}{l}{\it SAX1(1997)}\\ 
\hline

2 Gaussians& $6.4^f$&
$0^f$& $120^{+100}_{-80}$& $1.4\pm 1$& & 150/145\\
& $6.7^f$& $0^f$&  $350^{+160}_{-130}$& $3\pm 1$& & \\
\hline
\multicolumn{6}{l}{\it SAX2(1998)}\\ \hline
2 Gaussians& $6.4^f$& $0^f$& $80\pm30$&
$6.5\pm2.3$& & 391/372\\
& $6.96^f$& $0^f$& $60\pm30$& $4.6\pm2.1$& & \\
Fabian89& $6.4^f$& & $208\pm56$& $16\pm5$& $46\pm7$& 389/372\\
 Laor91& $6.4^f$& & $316\pm86$& $22\pm6$& $46\pm7$& 390/372\cr
\hline
\end{tabular}

$^f$Frozen value. Fabian89: disc line model for a
Schwarzschild black hole described by Fabian et al. (1989); Laor91: disc
line model for a Kerr black hole described by Laor (1991).
\end{table*}

Similar to the SAX1 observation, the iron line peak is higher than 6.4 keV
(although with a lower significance), while the line is definitely broad, 
($\sigma_{\rm K\alpha}=0.32^{+0.18}_{-0.06}$ keV) suggesting an
emission from a relativistic accretion disc (Fabian et al. 1989). The
confidence
contours between the line energy and width for the SAX2 data are shown in
Fig.~\ref{sax2_contour}. The line equivalent width is $EW_{\rm
K\alpha}=147\pm37$ eV, which may be easily explained by accretion disc
models where the matter is mildly ionized (Matt et al. 1992). 
The observed line
energy peak above the value of 6.4 keV expected from neutral matter could
be due to the Doppler shift in a relativistic disc inclined with
respect to the line of sight.\\ 

{\sl Disc line model}\\

Because of the above mentioned considerations, we
performed fits with line profiles derived from relativistic disc
models, under the assumptions of both a Schwarzschild and a Kerr metrics,
i.e. non rotating and rotating black hole, respectively (see
Tab.~\ref{irontab}).
Since the number of line parameters is relatively high and 
most of them would be poorly constrained in the fits, we fixed all
the line parameters but the inclination angle $\theta$ of the disc and
the line normalization. The line was assumed to be cold ($E_{\rm
K\alpha}=6.4$ keV). In the Schwarzschild metric we assumed for the
inner disc radius $R_i=6R_G$ (where $R_G=GM/c^2$ is the gravitational
radius and $M$ is the black hole mass), which corresponds to the last
stable orbit for an accretion disc
around a Schwarzschild black hole. The outer disc radius was fixed
to $R_o=1000R_G$ and the line emissivity was parameterized by $R^{-q}$,
where $R$ is the radius. The power law index $q$ was assumed equal
to 2.5, the mean value derived by Nandra et al. (1997b) for a sample of
Seyfert 1 galaxies in the case of a Schwarzschild black hole. In the Kerr
metric we assumed $R_i=1.23R_G$ (the last stable orbit in a rotating
black hole metric), $R_o=400R_G$, and $q=2.8$, the average value derived
from Nandra et al. (1997b) for a Kerr black hole. Fitting the iron line
with  
relativistic disc models gives $\chi^2/dof=389/372$ for the Schwarzschild
metric and $\chi^2/dof=390/372$ for the Kerr metric. Both
models provide an adequate description of the data, comparable with the
analytical broad Gaussian model. Remarkably, the inclination
angle of the disc turns out to be the same both from the fits with
the Schwarzschild and the Kerr metric: $\theta=46^o\pm7$. Therefore,
the disc seems to be inclined and the energy of the line centroid can
be explained by the relativistic Doppler shift of a cold line. 

The estimate of disc inclination depends on the choice of 
inner radius and so we tested different values of $R_i$ for the
Schwarzschild and the Kerr metric. In both cases, the best fit is found
when the inner radius correspond to the last stable orbit.  
The 90\% upper limits on $R_i$ are 
$30R_G$ and $5.3R_G$ for the Schwarzschild and the Kerr metric,
respectively. If we calculate the values of $\theta$ when $R_i$
is fixed to its 90\% upper limit, we find $\theta=65^{+25}_{-21}$ for the
Schwarzschild metric and $\theta=44^{+4}_{-5}$ for the Kerr metric.  
Also, if the disc is ionized, the line centroid could be higher than
6.4 keV. Then, the Doppler boosting effect could be lower and the
inclination estimate could be incorrect. We have therefore performed disc
line fits fixing the line energy peak at 6.62 keV, i.e. the SAX1 line
energy. The resulting values for $\theta$ are $36^{o\,+7}_{~-10}$ and
$38^{o\,+6}_{~-7}$ for a Schwarzschild and a Kerr black hole,
respectively,
fully consistent with $46^o\pm7$ derived previously for neutral disc
models.\\

{\sl Line blend model}\\

An alternate representation of the broad iron line feature could be a
blend of lines emitted by iron at different ionization stages.
We performed a fit to the iron complex with two narrow Gaussians
at 6.4 keV and 6.96 keV as expected from FeI--XVI 
and FeXXVI, respectively. This model provides an equally good
fit to the data as for the broad Gaussian line fit ($\chi^2/dof=391/372$).
Adding a 6.96 keV line describes the data better than having only the Fe
K$\beta$ component at 7.06 keV. The latter is not sufficient to
explain the data at $\sim 7$ keV.
The addition of a third line at 6.68 keV expected from FeXXV is not
significant. 

\subsection{Comparison with previous results}

\begin{figure}
\epsfig{file=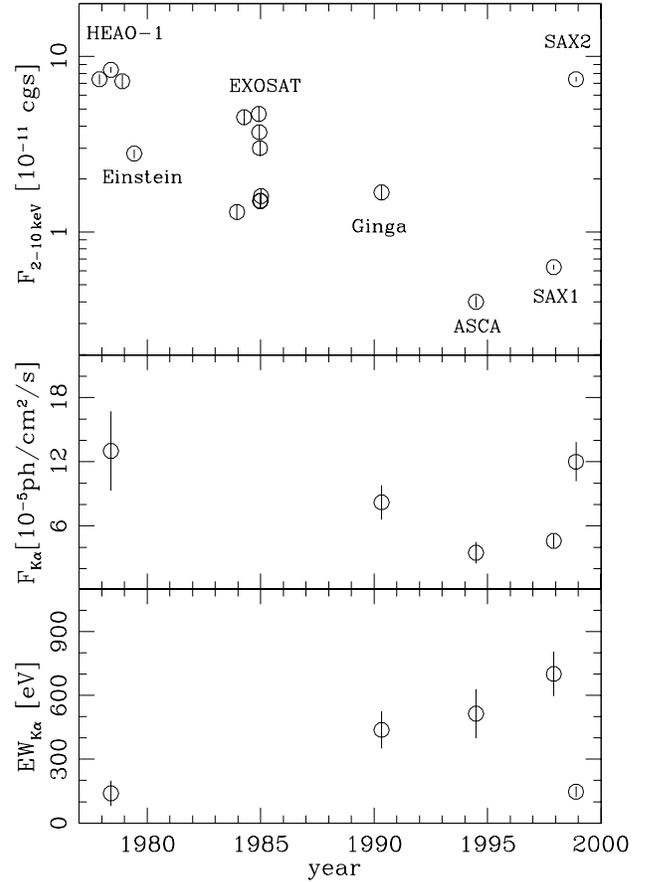, width=13.cm, height=13.cm}
\caption{From top to bottom, X--ray variability curve for the 2--10 keV
continuum flux, the iron K$\alpha$ line intensity, and the line
equivalent width. The HEAO--1, Einstein, EXOSAT, Ginga and ASCA data
are taken respectively from Mushotzky (1982), Turner et al. (1991), Turner
\& Pounds (1989), Nandra \& Pounds (1994), Weaver et al. (1996).
Errorbars are at the $1\sigma$ level.}
\label{variab}
\end{figure}

Before the BeppoSAX observations, the X--ray flux of NGC~2992
decreased
quite systematically from 1978 to 1994, when it was observed by ASCA. The
flux measured in SAX1 shows a modest increase with respect to the
ASCA value, while a factor of $\sim12$ increase is detected 
between SAX1 and SAX2. The 2--10 keV flux measured by SAX2
is similar to that measured by HEAO--1 in 1978. The long term
flux variability is shown in Fig.~\ref{variab}. The 
spectrum of NGC~2992 measured by SAX1 and SAX2 is an absorbed power
law with $\Gamma\sim1.7$ and $N_{\rm H}=10^{22}$ cm$^{-2}$ and is
on average consistent with the results previously obtained by HEAO--1,
Einstein, EXOSAT and Ginga, but not with the very flat
($\Gamma\sim1.2$) ASCA spectrum in 1994. The peculiar spectrum detected
by ASCA
has been interpreted by We96 as evidence for a strong reflection
continuum superposed to a weak power law continuum with $\Gamma\sim1.7$.
The reflection continuum should represent the echo of the past brightness
of the source. No significant evidence for short term variability was
observed in the ASCA data (We96).

An example of a similar flux and spectral variation has
been observed in NGC~4051 (Guainazzi et al. 1998), where a drop out of a
factor of $\sim 20$ in the 2--10 keV flux was accompanied by a spectral
change from a power law with $\Gamma\sim1.9$ to a pure reflection
continuum. 

The iron line in the BeppoSAX spectrum is
significantly different from that detected by ASCA. We96 found that 
the line in the ASCA spectrum is
cold ($E_{\rm K\alpha}=6.42\pm0.04$ keV), narrow ($\sigma_{\rm
K\alpha}<0.06$ keV) and has an
equivalent width of $EW_{\rm K\alpha}=514\pm190$ eV. Because of the line
narrowness and
its high equivalent width, We96 argued in favor of
iron emission from a distant torus, that must be characterized by a large
column density to produce the observed iron line and the cold reflection
that flattens the ASCA spectrum.
The line energy observed by Ginga in 1990 (Nandra \& Pounds 1994)
is $6.43\pm0.16$ keV, consistent with neutral iron emission. Because of
the
poor energy resolution of the Ginga LAC the line width was not determined.
No useful information about the line energy and width can be extracted by
HEAO--1 data (Weaver, Arnaud \& Mushotzky 1995).  
A summary of the variations of the iron line parameters
along all the X--ray observations of NGC~2992 is shown in
in Fig.~\ref{variab}.

\section{Optical and near--IR data}

As soon as we received the last set of BeppoSAX data and realized the
enhanced X--ray activity with respect to the previous BeppoSAX
observation,
we submitted a target of opportunity proposal to the European Southern
Observatory to observe the behavior of various nuclear components such as
the broad line region and the hot dust emission. As discussed in this
Section the observation were performed in early January 1999, i.e. about
1.5 months after the BeppoSAX observation.

\subsection{New observations and data reduction}

Near Infrared and optical observations of NGC~2992 were performed at the
ESO New Technology Telescope (NTT).

The infrared data were collected on January 7, 1999
with photometric conditions using the imaging and spectroscopic modes
of SOFI, equipped with a Hawaii HgCdTe 1024x1024 detector.

Images in the J, H and Ks filters were obtained
with the small field objective yielding a 0\farcs144/pix scale.
The seeing during the observations was $\sim 0\farcs7$.
Observations consisted of several on-source exposures interleaved with sky
exposures while stepping the object on the detector by $\sim 20\arcsec$.
The total on-source integration time was of 200~s in J and 120~s
in H and Ks.
Flat-field correction was performed using differential dome
exposures and the sky was subtracted from each on-source frame.
Images were then co-aligned by using stars present in the field of view
and coadded.
Flux calibration was performed using the standard star SJ9138 (Persson
et al. 1998).

\begin{figure}
\epsfig{file=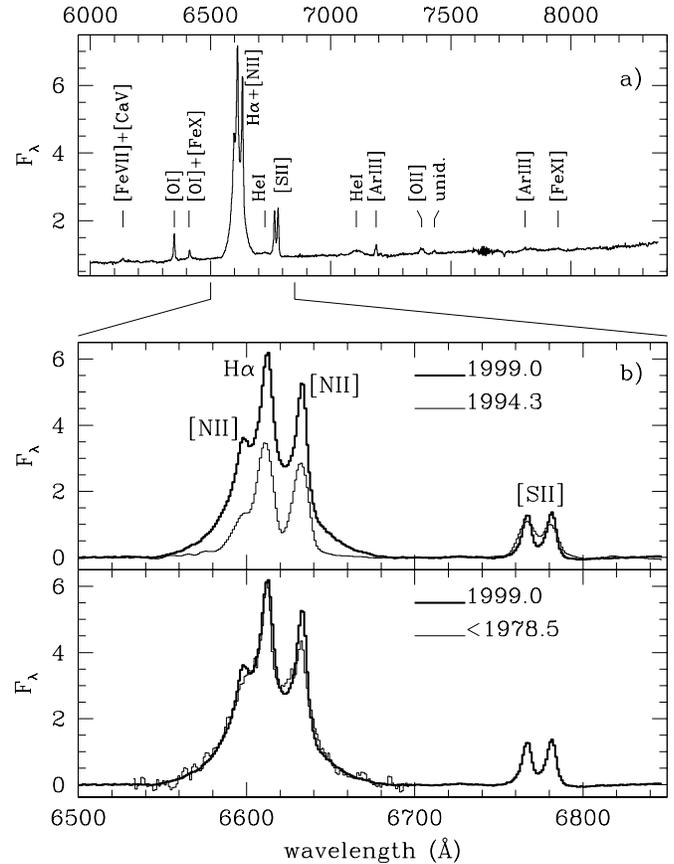, width=8.7cm, angle=0}
\caption{Top, optical spectrum of NGC~2992 obtained by us in January 1999.
Middle, comparison of our spectrum with that obtained by Allen
et al. (1999) in 1994, i.e. at the time of the minimum X--ray activity.
Bottom, comparison of our spectrum with that obtained by Veron et al.
(1980)
prior to 1978, i.e. when the nucleus was in the former high X--ray state.
}\label{opt_spectra}
\end{figure}

Spectral observations were performed with the Red and Blue grism
yielding a dispersion of 10 \AA/pix in the 1.52--2.52 $\mu$m 
range and 7 \AA/pix in the 0.95--1.64 $\mu$m range.
The pixels subtended 0\farcs276 along the 1\arcsec\ slit
and the resolving power
was $\sim 600$ at 2.2 $\mu$m and $\sim 500$ at 1.25 $\mu$m.
Observations consisted of several couples of exposures in which the
object was moved at two different positions along the slit, in order
to perform sky subtraction. On--chip integration time was 120 s resulting
in a total, on-source integration time of 960~s. 
Data were flat fielded with spectroscopic dome exposures and sky 
subtracted. 
Correction for optical distorsion of the slit direction and
wavelength calibration were performed with Xenon and Neon 
arc exposures. The zero point of the wavelength
calibration was then checked using the OH sky lines
(Oliva \& Origlia 1992).
Residual sky emission was removed by fitting a linear polynomium along the
slit.
Correction for instrumental response and telluric absorption, and
flux calibration were performed by using the G1V star HR3916.
The intrinsic stellar features were then canceled by dividing for the 
sun spectrum according to the
prescription described by Maiolino, Rieke \& Rieke (1995).
The FWHM (1\arcsec) of the point spread function along the slit
was estimated by using the wings of the broad lines
which are emitted from a spatially unresolved region.
The final spectrum was then extracted in a $1\arcsec\times 2\arcsec$
aperture centered on the nucleus which corresponds to the slit width
by 2xFWHM.

The optical spectra were collected on January 12, 1999
with photometric conditions using 
EMMI in the Low Resolution Spectroscopy (RILD)
mode with grating $\sharp6$.
The projected pixel size of the Tektronic 2048x2048 CCD
was 0\farcs27 along the slit and 1.2 \AA/pix along dispersion.
The slit width was 2\arcsec\
yielding a resolution of $R\sim 700$ at 6500~\AA.
The slit was centered on the galaxy nucleus with a position angle of
120 deg.
The 6000-8400~\AA\ spectrum was obtained by merging two different
exposures of 1200 s in order to remove cosmic rays.
Data reduction was performed using standard procedures
for CCD long--slit spectra, i.e. bias subtraction, flat field correction, 
rectification of the slit direction and wavelength calibration.
Correction for instrumental response and flux calibration were
performed using the spectrophotometric standard star LTT3864 (Hamuy et
al. 1992, 1994).
The final spectrum of the nuclear region was extracted, 
with the same procedure adopted for the IR spectra, 
in a $2\arcsec\times1\farcs4$ aperture.

\subsection{Broad lines analysis}

\begin{table*}
\caption{Properties of the optical and near infrared hydrogen lines.}
\label{hydrotab}
\begin{tabular}{ccl}
\hline \hline
 & 2190 & (1999.0)$^b$ \\
FWHM[$H\alpha(broad)$]$^a$ & 2620 & (1994. - near minimum)$^c$\\ 
 & 2030 & ($<$1978.5 - former active phase)$^d$ \\
\hline
 & 6.8 & (1999.0)$^b$ \\
$\displaystyle
H\alpha(broad) \over
\displaystyle H\alpha(narrow)$ & 0.7 & (1994. - near minimum)$^c$\\ 
 & 6.0 & ($<$1978.5 - former active phase)$^d$ \\
\hline
 & 17.2 & (1999.0)$^b$ \\
$\displaystyle
Pa\beta(broad) \over
\displaystyle Pa\beta(narrow)$ & 1.7 & (1992.4 - near minimum)$^e$\\ 
 & 3.3 & (1989. - intermediate phase)$^f$ \\
\hline
$\displaystyle
Br\gamma \over
\displaystyle Pa\beta $ & 0.22 & (1999.0)$^b$\\ 
\hline
$\displaystyle
Pa\beta (broad) \over
\displaystyle H\alpha (broad)$ & 0.30 & (1999.0)$^b$\\ 
\hline \hline
\end{tabular}
\\
$^a$ Full width half maximum in km/s.\\ $^b$ This work.\\
$^c$ Allen et al. (1999).\\ $^d$ Veron et al. (1980).\\
$^e$ Goodrich et al. (1994).\\ $^f$ Rix et al. (1990).
\end{table*}

\begin{table*}
\caption{Near infrared photometry of the nucleus.}
\label{irphottab}
\begin{tabular}{ccccccc}
\hline \hline
Epoch & Aperture & J & H & K & J--K & H--K \\
\hline
1999.0$^a$ & 2$''$ & 12.52 & 11.50 & 10.63 & 1.89 & 0.87 \\
1990.0$^a$ & 12$''$ & 10.94 & 10.01 & 9.45 & 1.49 & 0.56 \\
1988.3 (burst)$^b$ & 12$''$ & 11.00 & 9.93 & 9.21 & 1.79 & 0.72 \\
1987.6 (quiesc.)$^b$ & 12$''$ & 11.18 & 10.21 & 9.77 & 1.41 & 0.44 \\
\hline \hline
\end{tabular}
\\
$^a$ This work.\\
 $^b$ Glass (1997)
\end{table*}

\begin{figure*}
\epsfig{file=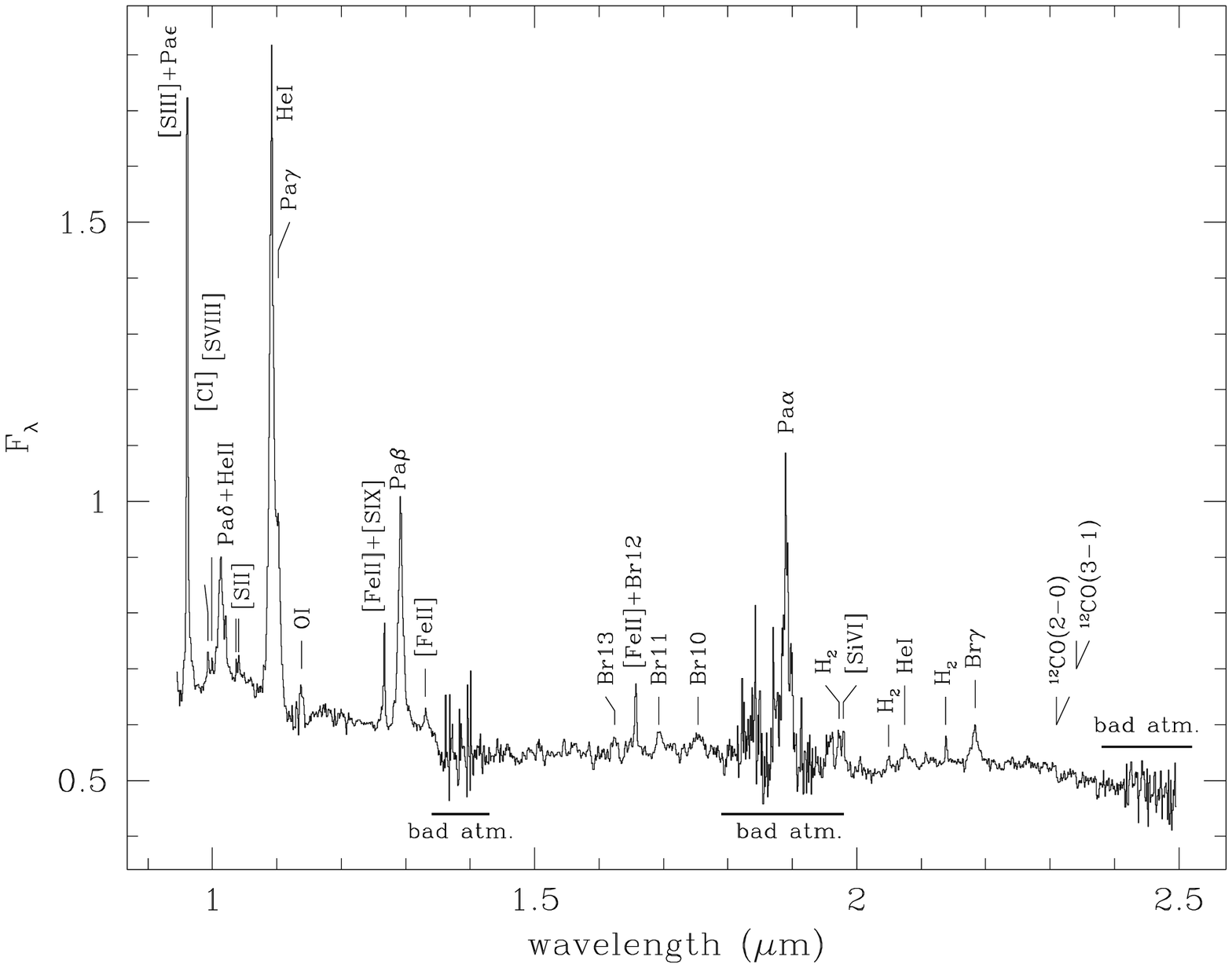, width=14cm,
angle=0}
\caption{Near infrared spectrum of NGC~2992 obtained in January 1999.
}\label{ir_spectrum}
\end{figure*}

Fig.~\ref{opt_spectra}a shows the optical spectrum of NGC~2992 obtained
with EMMI.
Fig.~\ref{opt_spectra}b shows the continuum subtracted zoom of the
spectrum around H$\alpha$, that
clearly shows the presence of a prominent broad component of H$\alpha$.
For comparison Fig.~\ref{opt_spectra}b
also shows the nuclear spectrum obtained by Allen
et al. (1999) in March 1994, i.e. at the minimum activity ever observed in
the
X--rays. The latter spectrum was scaled to match the intensity of the
[SII]
doublet in our spectrum. Although Allen's spectra have slightly lower
spectral
resolution ($\sim 550$) their spectrum clearly shows a much weaker, if
any,
broad component of H$\alpha$. In contrast, our spectrum is
remarkably similar to the spectrum obtained by Veron et al. (1980) prior
to
1979\footnote{The date of the observation is not given in the Veron et al.
paper.}, i.e. probably when the nucleus was at the same level of
X--ray activity as in late 1998 (Sect.~\ref{xdata}). We deblended the
broad
and
narrow
components of H$\alpha$ by using a multiple Gaussian fit. The narrow   
components are unresolved at our spectral resolution.
Some relevant results of the fit are reported in
Tab.~\ref{hydrotab} (additional information is given in the Appendix) and
are compared
with
the values obtained by
Veron et al. (1980) in the former high activity phase.
The ratio between broad and narrow
H$\alpha$ are nearly identical for the two observations. Also
Allen et al. spectrum
seems to require a broad H$\alpha$ component of about the same width as
that found in our spectrum, but about 10 times weaker.

The optical spectrum also
shows a variety of other lines that are marked in Fig.~\ref{opt_spectra}a.
The analysis of flux  
variation of some coronal lines (such as [FeVII]$\lambda$6087\AA, 
[FeX]$\lambda$6374\AA, [FeXI]$\lambda$7892\AA\ and 
[SiVI]$\lambda$1.96$\mu$m)
with respect to the past would provide interesting constraints
on the size of the Coronal Line Region, that is expected to be
intermediate
between the BLR and the NLR. Unfortunately, the quality of previous
spectra
is too poor to provide useful constraints to the flux of such coronal
lines.
However, in the Appendix we provide the fluxes of all the lines that are
detected and identified in our spectrum and that could be useful for
comparison with future monitoring.

Fig.~\ref{ir_spectrum} shows the infrared spectrum taken with SOFI at
ESO, where we
combined the blue and red spectra. The spectrum shows a wealth of
features,
most of which are observed for the first time in this galaxy thanks to the
much improved sensitivity of this observation with respect to previous data.
In this paper we only discuss those features that are relevant to the
issue of the burst observed in the X--rays. However, in the Appendix we
report the list of the lines, along with their fluxes,
that we detected and identified in the spectrum. Pa$\beta$ is the
most suitable line for studying the profile of near--IR hydrogen lines,
since it
is relatively strong, isolated, and in a good atmospheric window. We fitted the
Pa$\beta$ with two Gaussians (broad and narrow, unresolved)
whose parameters were frozen to the same values
obtained for H$\alpha$ except for their normalization. As reported in
Tab.~\ref{hydrotab}
the observed Pa$\beta$ can be almost fully accounted for by the broad
component alone, while the narrow component contributes for only $\sim$7\%,
much lower than in the case of H$\alpha$. In Tab.~\ref{hydrotab} we also
compare our
result with those obtained by Rix et al. (1990) in 1989, when the nucleus was
mildly active, and with the result of Goodrich et al. (1994) in May
1992, i.e.
when the nucleus was close to its minimum activity. Although this comparison
should be considered with caution because of different instrumental setup
and lower sensitivity of past observations, there is an apparent correlation
between the Pa$\beta$(broad)/Pa$\beta$(narrow) ratio
and the level of nuclear activity as traced by the hard X--rays.

\subsection{The hot dust emission}

The most interesting feature of the near infrared images (not shown)
 is a powerful
nuclear emission characterized by very red colors. More specifically,
within the central 2 arcseconds we measure $\rm H-K = 0.9$ and $\rm J-K =1.9$,
while typical galaxy colors are $\rm H-K = 0.2$ and $\rm J-K \simeq 0.6$.
Nuclear dust absorption is probably partly responsible for such red colors, but
very likely hot dust emission in the H
and K bands also contributes significantly
(see Oliva et al. 1995, Maiolino et al. 1998, Thatte et al. 1997 for similar
cases in other objects). This
interpretation is supported, in the specific case of our observations, by the
near--IR spectrum: the stellar CO absorption bands above 2.29 $\mu$m
and in the H band (e.g. at 1.62 $\mu$m) are much weaker than generally
observed in
normal galaxies. In particular, the equivalent width of the
$\rm ^{12}CO(2-0)$ band is only 7.5 \AA, while in normal galaxies this
quantity generally ranges between 12 and 14 \AA. This effect has been 
observed in other AGNs and is generally ascribed to hot dust emission
diluting the stellar features in the near--IR (Oliva et al. 1995, Oliva et
al. 1999).

The burst observed by Glass (1997) in 1988 during his near--IR monitoring
also had features typical of hot dust emission, indeed the burst had much
higher power in the K and L bands than in H and J. In Tab. 5 we compare
the magnitudes observed by us
with the magnitudes measured by Glass at the maximum, in the same aperture,
during the 1988 burst,
and the background galaxy magnitudes that corresponds to 
the minimum values observed during the near--IR
monitoring.\footnote{With regard to the K band magnitudes, note that we
used a filter (K-short) slightly different from that used by Glass, that is
less sensitive to the hot dust emission at long wavelengths.}
The magnitudes measured by us are intermediate between the maximum and the
minimum value observed. The physical implications are discussed in the next
Section.

\section{Discussion}\label{discussion} 

\subsection{Long and short term variability in the X--rays}

The data collected by BeppoSAX suggest that an AGN has revived in
the center of NGC~2992. The 2--10 keV flux of the source 
systematically decreased by a factor of 19--20 between 1978 and 1994
and then increased by the same amount from 1994 to 1998. 
During the same time intervals the iron line flux changed by a factor
$\sim 3.5$. Therefore, in 1998 both the continuum and the iron 
K$\alpha$ line intensity have settled back to the same values observed 
in 1978. This coincidence of fluxes could indicate that in the high state the 
luminosity of NGC~2992 is close to the maximum luminosity achievable, 
i.e. its Eddington limit. However further observations would be needed to 
confirm this hypothesis.

The rebuilding of the AGN is also supported by the analysis of the short
term variability of NGC~2992.
In 1998, during the high state SAX2 observation, the source
exhibited significant flux variations with an amplitude typical of Seyfert
1s (Nandra et al. 1997a), while in the low states observed in 1994 and
1997 no evidence for short term variability was found.
The SAX2 variability is consistent with a
scenario where we are observing the AGN directly through an
absorbing medium. This situation is frequently
observed in other objects with an intermediate Seyfert type classification
in the optical and a column density of $N_{\rm H}\sim 10^{22}$ cm$^{-2}$
in the X--rays (e.g. NGC~526A, NGC~7314 and MCG~--5-23-16; Turner et al.
1997).  

The lack of variability thoughout the SAX1 observation could suggest that
during the rebuilding phase of the AGN the geometry of the central region
and the accretion processes were different from those observed in SAX2.
Ptak et al. (1998) found that Low Luminosity AGNs (LLAGNs) do not follow
the Nandra et al. (1997a) relation, i.e. LLAGNs do not show time
variability. These authors ascribed the behaviour of LLAGNs to the
presence of an Advection Dominated Accretion Flow (ADAF; see Narayan,
Mahadevan \& Quataert 1998 for a review) in their nuclei.
We note, however, that the X-ray luminosity of NGC
2992 during the SAX1 observation ($1.8\times10^{42}$ erg s$^{-1}$) is
rather high compared with the typical luminosity of LLAGNs
($10^{40}-10^{41}$ erg s$^{-1}$). 
Furthermore, the ADAF scenario would hardly explain the prominent iron line 
detected in SAX1.
Therefore, we favor the interpretation that ascribes the lack of
variability in SAX1 to a distant and extended forming disc (and, possibly,
an extended corona) rather than to an ADAF. Another possibility is that
during the 
SAX1 observation the nucleus was in a transition state where both a
central ADAF and a
distant accretion disc were present.  

Since the long term variations amplitude of the iron line flux is
different from that of the continuum, it is very likely that an accretion
disc is not the only source of the line photons. Indeed, the disc is 
close to the X--ray primary source and it should
reprocess the primary radiation on very short timescales of minutes or
hours. Thus, one should expect long term variations of the
disc line flux identical to those of the continuum. 
An iron emission from a
distant medium, such as the obscuring torus postulated by the unified model,
which lags the direct flux by timescales of
years is therefore very plausible. We96 derived that the torus must have a
geometry and distance from the nuclear source such that it produces a time
lag of $\sim 10$ yr in the variation of the Fe line with respect to the
continuum.
Although this is likely to be the case for the narrow line
detected by ASCA, no constraints are present on the line width
detected by HEAO--1 in 1978. We note that if a disc component is present
in
the HEAO--1 data, the estimate of We96 is likely to be a
lower limit on the torus distance.

\subsection{Iron line and reflection continuum diagnostics in SAX1}

The SAX1 observation could have caught NGC~2992 during the rebuilding  
phase of the AGN. The spectral continuum is similar to that of normal
AGNs, while the iron feature is peculiar and cannot be easily explained in
the framework of iron emission from an accretion disc or a molecular
torus (Ghisellini, Haardt \& Matt 1994). Indeed, the line is narrow 
in contrast with standard disc models. Furthermore, the line energy
($E_{\rm K\alpha}=6.62\pm0.07$ keV) is not consistent with 6.4 keV as
expected from the neutral iron in the torus.

Such a narrow and warm line could be produced by ionized iron in a
warm scattering material via resonant scattering (Matt, Brandt
\& Fabian 1996). A problem with this model is related to the observed
line equivalent width of $\sim 700$ eV.
The equivalent width of the resonant lines predicted by
Matt et al. (1996) can be as high as a few keV if the nuclear
source is obscured and only the scattered continuum observed. 
 This is the case of NGC 1068 (Ueno et al. 1994; Matt et al. 1997) 
which is a Compton thick AGN. However, when the nuclear source is directly
viewed, as in the case of NGC~2992, the equivalent width of the resonant 
lines is predicted to be at most $\sim 100$ eV (Krolik \& Kallman 1987).

Another possible explanation of the line properties in SAX1 is to assume a
blending between a narrow cold component from the molecular torus and a
narrow hot component from a distant ionized disc which is forming.
The narrowness of the line can be explained by
assuming that the accreting material which is going to rebuild the disc is
still far from the central black hole and so the line does not suffer
from strong Doppler or gravitational broadening. 
We have therefore fitted the SAX1 line data with two narrow Gaussians at
6.4 keV and 6.7 keV, also adding a reflection continuum. 
In order to compare the
intensity of the 6.4 keV line and of the reflection continuum with the
ASCA values, we have fixed the spectral index
to $\Gamma=1.7$, as adopted by We96 for the ASCA data.
We obtain a fit as good as the single Gaussian
line fit; the equivalent widths and intensities of the two lines
are respectively: $A_{6.4 \rm{keV}}=1.4(\pm 1)\times10^{-5}$ photons
cm$^{-2}$ s$^{-1}$, $EW_{6.4 \rm{keV}}=120^{+100}_{-90}$ eV, $A_{6.7
\rm{keV}}=3(\pm 1)\times10^{-5}$ photons
cm$^{-2}$ s$^{-1}$, $EW_{6.7 \rm{keV}}=350^{+160}_{-130}$ eV. 
The 6.4 keV line intensity is lower than the ASCA value, qualitatively 
in agreement with the idea that this is a fading reprocessed component 
(quantitatively, the fading of the Fe cold line depends on the detailed 
geometry of the torus).
The comparison between the ASCA and SAX1 reflection continua is less
straightforward since a disc contribution is expected in SAX1 in addition
to the torus one. However, we note that in SAX1 data there is scope for
including a torus reflection normalization lower than the ASCA value.

Both the centroid energy and equivalent
width of the 6.7 keV component could be explained by means of ionized disc
models (Matt, Fabian \& Ross 1993). The ionization structure of the disc
is basically determined by the ionization parameter $\xi$
defined as following: 

\begin{equation}\label{xi}
\xi=4\pi F_{\rm h}/n_{\rm H}, 
\end{equation}

where $n_{\rm H}$ is the hydrogen number density of the
disc and $F_{\rm h}$ is the hard X--ray flux impinging on the disc
surface. When $\xi$ is of the order of $\sim$ a few hundreds (in units
of erg cm s$^{-1}$), the iron ionization state
is mostly below FeXVI: the line
energy is at $\sim 6.4$ keV and the equivalent width is $\sim 150$ eV as
expected for cold discs. For $\xi\sim2000$ the emission is
dominated by FeXXV: the line energy is close to 6.7 keV and the
equivalent width is as high as 350--400 eV since most of the elements
lighter than iron are fully stripped and do not absorb the iron line
photons. Finally, when $\xi\sim5000-10000$ the bulk of the line is
emitted by FeXXVI: the line energy is at 6.96 keV and the
equivalent width has decreased to $\sim 100$ eV, since a large fraction
of iron is now completely bare. Therefore, the 6.7 keV line in the SAX1
spectrum is consistent with emission from a ionized disc with
$\xi\sim2000$. Following Matt et al. (1992), if we assume that the 
hard X--ray source (possibly an extended corona of relativistic electrons, 
Haardt \& Maraschi 1991) is located at $\sim 50R_G$ above the disc plane, 
we can estimate a hydrogen density of $n_{\rm H}\sim 10^{14}$ cm$^{-3}$ 
for these stages of the rebuilding disc.

\subsection{Iron line and reflection continuum diagnostics in SAX2}

The SAX2 spectrum of the high state is
consistent with the typical X--ray spectrum of AGNs, suggesting that the 
rebuilding of the AGN phenomenon is almost completed. The iron line
profile in SAX2 is consistent with that expected from cold accretion disc
models. The hard X--ray
luminosity has increased by an order of magnitude with respect to the SAX1
data, and $\xi$ is expected to be of the order of $\sim 200$ as for cold
discs. Therefore, the full-developed accretion disc in
SAX2 should be about two order of magnitude denser than the forming disc
observed in SAX1. If we assume that at the time of the ASCA
observation the AGN was in a completely quiescent phase, we can consider
the time interval between the ASCA and SAX2 observations as un upper
limit to the timescale for the rebuilding of the disc. Also, the time
interval between SAX1 and SAX2 provide a lower limit to the
rebuilding timescale. As a result, the rebuilding timescale turns out to
range from 1 to 5 years.
Since the obscuring torus should reside at a distance of
$\sim 3.2$ pc (i.e. 10 ly) from the central source (We96), in SAX2 (1998)
any narrow iron line from the torus, in addition to the disc line, is
expected to be weaker than that detected by We96 in 1994. We have
verified this to be the case. Also, the reflection component produced by
the torus is now completely diluted by the nuclear and disc continua.
The best fit inclination angle for the disc, as
estimated from the iron line profile, is $\theta=46^o\pm7$.
Provided that the accretion disc and the molecular torus are coplanar, the
inclination angle measured from the disc line profile is in agreement with
the unified schemes where intermediate Seyfert types could be active
nuclei observed through the rim of the torus. The moderate X--ray
absorption of $N_{\rm H}\sim 10^{22}$ cm$^{-2}$ could be indeed
produced by the outer boundary of an optically thick torus. 
Alternately, the X--ray absorption could be produced by the gas
associated to the dust lane observed in the host galaxy and crossing the
nucleus. In line with this interpretation is the fact that the optical
extinction inferred from the optical and infrared
{\it narrow} lines is close to that expected from the gaseous
column density derived from the X--rays (for a Galactic gas-to-dust
ratio).
The fact that
that the extinction measured toward the BLR is similar to the extinction
affecting the NLR is also in favor of this intepretation.
The nature of the absorber will be discussed more in detail in the next
section.

We note that NGC~2992 could be one of the first sources where there is
evidence for two distinct absorbers in different spatial regions: the
first (along our line of sight)
is associated to the interstellar gas
extended on the 100 pc scale, and the second (out of the line of sight) is
associated to the thick molecular
torus on parsec scales.
This phenomenology
has been recently observed also in the Seyfert 1.8 galaxy NGC~1365
(Risaliti, Maiolino \& Bassani 1999).

In SAX2 one problem related to the AGN scenario with an accretion disc
model is that the reflection continuum produced by the disc and associated
to the iron line
is weaker than expected. The relative normalization between the direct and
reflected continuum is $R=0.12$, which is in contrast with the value of 1
predicted by an isotropic source illuminating a plane--parallel
semi--infinite
slab and suggested by the line equivalent width. If we introduce 
a high--energy spectral cut off to $E_c=150$ keV, the minimum e-folding
energy value to maintain a statistically good fit, then 
$R$ increases only to 0.4, which is still a low value.
Two possibilities to
solve this problem are to introduce an iron overabundance or
to assume that we are observing the source through a dual absorber with
column densities of $\sim10^{22-23}$ cm$^{-2}$. 
The combination of a dual absorber, a 
canonical power law with $\Gamma\sim1.9$, and a reflection component could
indeed mimic a flatter power law with $\Gamma\sim1.7$. In this scenario
the line photons could be produced both in reflection by the accretion
disc and in transmission by the dual absorber. Examples of complex
absorbers are found in the X--ray spectrum of NGC~2110 (Malaguti et al.
1999) and NGC~5252 (Cappi et al. 1996). 

In Section~\ref{xdata} we showed that another representation of the iron
line complex of SAX2 could be a blending between lines emitted by
different iron ions. In particular, two narrow lines at 6.4 keV and at
6.96 keV, deriving from the low and highly ionized iron respectively, 
adequately fit the data. However, this scenario seem to be
disfavoured since the intensity of the cold line at 6.4 keV is higher than
expected. Indeed, the narrow cold line at 6.4 keV in SAX2 should be
the signature of the emission from the torus and, therefore, should
be weaker than that measured by ASCA, just because the nuclear continuum
had been falling until 1994 and there is a 10 years time lag between
narrow-cold Fe line and continuum (We96).
Actually, the 90\% lower limit to the intensity of the narrow cold
line measured in SAX2 (see Table~\ref{irontab}) is higher than the 90\%
upper limit to the ASCA line intensity ($3.5\times 10^{-5}$ photons
cm$^{-2}$ s$^{-1}$) derived by Turner et al. (1997).

\subsection{The BLR and the hot dust emission}

Unfortunately, the optical and near-infrared follow up performed by us
cannot
provide tight constraints to the size of the Broad Line Region and
of the hot dust emitting region because we do not know exactly the time
of the X--ray burst; we only know that the transition
occurred between December 1997
and November 1998. Nonetheless our data provide some interesting information
on the geometry of the circumnuclear region.

Reverberation studies have shown that the BLR has typical sizes
$\rm R_{BLR} = 0.02 L_{45}^{1/2}~pc$ (Netzer 1990). NGC~2992 has a hard
X--ray
luminosity of $\rm L_{2-10keV}=2\times 10^{43}~erg~s^{-1}$, that translates
into a bolometric luminosity of about $\rm L_{bol}=2\times
10^{44}~erg~s^{-1}$ (Mulchaey et al. 1994). Therefore the expected size
of the BLR is about 0.01 pc. Our detection of broad lines both in the optical
and in the near--IR spectra can only provide an upper limit to the size of
the
BLR, given by the fact that the burst cannot be older than about 1 yr at the
time of the optical-IR
observations and, therefore, $\rm R_{BLR}(NGC~2992) < 0.3~pc$
that is fully consistent with the expected value of 0.01 pc
derived above.
However, we note that reverberation studies are generally more
sensitive to the inner region of the BLR, since this is the region
where rapid line variations are better detected. Fig.~8 and Tab.~4
show that the
broad lines detected by us are at the same level as those observed by Veron
et al. (1980) during the former high level state prior to
1978. This implies
that the whole BLR has already been fully ionized and, therefore, the upper
limit of 0.3 pc does not simply apply to the average BLR radius or to
the ``cross-correlation'' size (that is usually the quantity measured by
reverberation mapping), but this upper limit applies also to the outer
radius of the BLR.

As discussed in the former Section both the IR spectrum and images clearly
show evidence for hot dust emission. Although dust close to the
sublimation limit (T$\sim$1500 K) has the highest emissivity at $\sim
2\,\mu$m, the increasing
volume of emitting dust at lower temperatures implies that a large contribution
to the K band emission comes also from regions at lower temperatures (down to
400 K), hence at larger distances. The K and H band fluxes observed by us
are not at the maximum level observed by Glass (1997) during the 1988
burst\footnote{This is not necessarily the maximum near--IR emission that
can be achieved by the nucleus, since in 1988 the burst was temporary and
did not last very long and, therefore, probably not all of the hot dust 
emitting region was completely formed.},
implying that in January 1999
the hot dust emission is still in the raising phase.
The region
of dust close to the sublimation limit (T$>$1000 K) was already heated. 
Instead, the regions with dust at temperatures around 500 K, that significantly
contribute to the K band emission, were not active yet. According to the dust
equilibrium temperature obtained by
Laor \& Draine (1993), dust at temperatures around 500 K should be located at
a radius of about
$\rm R_{pc} \simeq 0.36 L^{1/2}_{46} T^{-2.5}_{1500} = 0.8$ pc,
given the luminosity of NGC~2992, and therefore it
is not surprising that we do not see the whole K band emission yet,
considered that the burst was not older than 1 year at the time of the
infrared observations. The dust close to the sublimation limit 
($\rm T\simeq 1500$ K) is expected to be located at a distance of
$\sim 0.05$~pc, fully consistent with the upper limit of 0.3 pc that we
can infer from the maximum time delay between the X--ray burst and the
infrared
observation.

Generally, the sizes of BLR and 
hot dust region have been estimated independently for different objects.
However, Netzer \& Laor (1993) proposed that the relative dimensions of these
two regions are related and, more specifically, that the outer boundary of the
BLR is located just inside the sublimation radius of the dust grains. This
model was proposed to explain the apparent gap between the BLR and the NLR with
almost no line emission. Our optical and infrared followup supports this model.
Indeed, as discussed above, the BLR appears to be fully formed when
the hot dust region is still forming, consistent with the fact that
the BLR outer radius is smaller than the hot dust emitting region.

\subsection{Optical extinction}

Finally, we can use the ratio between the broad lines to estimate the
absorption toward the BLR and compare it with the absorbing column density
derived from our X--ray observations. As discussed in the previous
Section and as shown in Tab.~\ref{hydrotab}, in the near IR the hydrogen
lines are completely
dominated by the broad components. By assuming an intrinsic ratio 0.16
between Br$\gamma$ and Pa$\beta$, and the extinction curve in
Rieke \& Lebofsky (1985), we infer
$\rm A_V(BLR) = 2.2\pm 0.6 ~mag$. Instead, by comparing
Pa$\beta$ with the broad component of H$\alpha$, and assuming an
intrinsic Pa$\beta$/H$\alpha$ = 0.052, we derive $\rm A_V(BLR) = 3.5\pm
0.4~mag$.
However, the latter value is less reliable both because radiation transport
and collisional excitation might affect the intrinsic Pa$\beta$/H$\alpha$
ratio in the BLR, and because problems
of intercalibration between the optical and the infrared spectra might
affect the observed value.
The extinction toward the NLR can be derived by the narrow components
of Pa$\beta$ and H$\alpha$: we obtain
$\rm A_V(NLR) = 2.0\pm 0.6 ~mag$\footnote{Ward et al. (1980) obtained a
value of $\rm A_V(NLR) =4.2$
from the narrow lines H$\alpha$/H$\beta$ Balmer decrement.
Part to the difference with our result is to ascribe to a different extinction
curve and an adopted
intrinsic ratio lower than currently assumed for the narrow line
region of AGNs (i.e. 2.8 instead of 3.0); if we use their narrow line Balmer
decrement with our extinction curve and the ``correct'' intrinsic ratio
we obtain $\rm A_V(NLR)=3.7$. Moreover the H$\beta$ line is probably affected by
the corresponding stellar H$\beta$ in absorption 
(Ward et al. spectrum shows evidence for stellar
features) that artificially increases the Balmer decrement. Finally the
limited spectral resolution might have resulted in problems with the emission
lines deblending.}, consistent with what obtained for the
broad line region. This finding suggests that both BLR and NLR are
absorbed by the same (large scale) medium, possibly the 100 pc scale dust
lane observed in the images.

The optical-infrared absorption towards the BLR is lower than that 
expected from the $N_H$ measured from the X--ray spectrum. By assuming a 
Galactic gas-to-dust ratio we find $\rm A_V(X) = 4.5\times
10^{-22}N_H(cm^{-2}) = 4.5$.
As suggested by Granato et al. (1997), this mismatch between optical
and X--ray absorption might indicate the presence of a dust-free region in
the
X--ray absorber. Alternatively, the obscuring gas might be characterized
by a
gas--to-dust ratio lower than Galactic or the average dust grain
size might be larger than in the interstellar Galactic medium, thus
making the extinction curve flatter (Laor and Draine 1993).

\section{Conclusions}

In this paper we have shown that an AGN has revived in the center
of NGC~2992. The spectral
continuum detected both in SAX1 (1997) and SAX2 (1998) can be described
with a power law with $\Gamma=1.7$ absorbed by $N_{\rm H}=10^{22}$
cm$^{-2}$, which is consistent with the mean spectrum of NGC~2992 observed
in the past. A prominent iron K$\alpha$ line is also detected in both the 
data sets.
The SAX1 observation is located between the ASCA and the
SAX2 observations (which correspond to the lowest and
highest states of the source, respectively), and seems to have caught
the source during the rebuilding phase of the accretion disc. From the
properties of the iron line detected in SAX1, the forming disc appears to
be constituted by gas which is still far form the central
black hole and is highly ionized, possibly because of its low density.
During the SAX2 observation the disc appears to be almost completed and
the iron line properties fit into the standard (cold) disc models. 

We also presented the results of optical and near--IR followup
observations
obtained 1.5 months after the detection of the X--ray burst. The optical
and near--IR hydrogen lines are characterized by prominent broad
components,
with intensities comparable to those observed in the former active phase
prior to 1979. The near--IR images and the near--IR spectrum also show
evidence
for hot dust emission in the H and K bands. However, at the time of the
observations the K and H band nuclear
emission was still not as high as observed in a previous IR burst, suggesting
that hot dust region was still in the process of forming.

\begin{acknowledgements}
We thank Chris Lidman, the NTT and SOFI staff for performing the
near infrared and optical observations. We thank Roberto Della Ceca
for helpful comments. We are grateful to M. Allen and M. Dopita for
providing their optical spectrum of the nucleus of NGC~2992. 
The referee is thanked for his comments. This 
work was partly supported by the Italian Space Agency
(ASI) under grant ARS--98--116/22 and by the Italian Ministry for
University and Research (MURST) under grant Cofin98--02--32.
\end{acknowledgements}

\appendix
\section{List of line fluxes}
{\normalsize
In Tab. A1 we list the lines detected in our optical and near--IR
spectra. In the case of permitted lines, the flux given in the last column
gives the total line flux, i.e. both broad and narrow components. In the case
of blending the line fluxes were obtained by means of a multiple Gaussian
fit.
In cases for which the blending is excessive to allow a deconvolution we give
the total flux of the blend.

We could not derive a flux for the HeI $\lambda$6678 since it is severely
blended
with the H$\alpha$ wings and with the [SII] doublet, resulting in a plateau
between the two groups of lines. Some uncertainties exist for the
identification of the line at 13305 \AA\ : it could either be fluorescence
OI $\lambda$13170\AA\ or [FeII] $\lambda$13205\AA. We favor the latter 
identification
both because it matches much better the observed wavelength and because the 
observed
flux is in excellent agreement with that expected from [FeII]$\lambda$12567\AA\
(the observed ratio between the two lines is 3.26, while the theoretical
value is 3.75). The absence of OI $\lambda$13170\AA\ strongly suggests that the
OI $\lambda$11290\AA\ line is produced by {\it Ly$\beta$ fluorescence} and 
{\it not}
by continuum fluorescence (in the latter case both OI lines should have 
nearly the
same intensity). Ly$\beta$ fluorescence is also supported by the lack of OI 
lines
at 7002 \AA\ and 7255 \AA\ . We could not find any identification
for the line detected at 7431 \AA\ , possibly this might be an instrumental
artifact.}

\begin{table}[b]
\caption{Optical and near--infrared emission lines detected in our spectra.}
\label{opt_ir_lines}
\begin{tabular}{lcc}
\hline \hline
Identification & $\lambda_{\rm obs}^a$ &  Flux$^b$ \\
\hline
$[$OI$]\lambda$6300    &    6349.5  &    5.2 \\
$[$OI$]\lambda$6364    &    6413.6  &    1.74 \\
$[$FeX$]\lambda$6374   &    6421.   &    0.6$^c$ \\
$[$NII$]\lambda$6548   &    6597.8  &   8.0 \\
H$\alpha$~$\lambda$6563   &    6612.   &   195. \\
$[$NII$]\lambda$6583   &    6633.5  &   24. \\
HeI $\lambda$6678     &    6728.   &   --$^f$ \\
$[$SII$]$$\lambda$6716   &    6767.   &   11.1 \\
$[$SII$]$$\lambda$6731   &    6781.   &   11.9 \\
HeI$\lambda$7065     &    7110.   &    7.8 \\
$[$ArIII$]$$\lambda$7136 &    7188.   &    2.7 \\
$[$OII$]$$\lambda$$\lambda$7318-7330 & 7379.   &    3.3 \\
unident.      &  7431.   &    0.95 \\
$[$ArIII$]$$\lambda$7751   &  7810.   &    0.34 \\
$[$FeXI$]$$\lambda$7891    &  7948.   &    0.12 \\
$[$SIII$]$$\lambda$9531    &  9605.1  &    31. \\
Pa$\epsilon$~$\lambda$9546    &  9619.5  &    26.$^c$ \\
$[$CI$]$$\lambda$9850      &  9930.   &     2.2$^c$ \\
$[$SVIII$]$$\lambda$9912   &  9990.   &     2.2$^c$ \\
Pa$\delta$~$\lambda$10049 & 10126.   &   26.5 \\
HeII$\lambda$10123     & 10201.   &     3.7$^c$ \\
$[$SII$]$$\lambda$10286    & 10365.   &     2.2$^c$\\
$[$SII$]$$\lambda$$\lambda$10320+10336 & 10408. &     2.9$^c$ \\
HeI$\lambda$10830      &  10917.  &     123.6 \\
Pa$\gamma$~$\lambda$10938   &  11023.  &     37.$^c$ \\
OI$\lambda$11290       &  11383.  &     6.9$^d$ \\
$[$SIX$]$$\lambda$12524& 12622 &   1.8$^c$ \\
$[$FeII$]$$\lambda$12567   &  12669    &    7.5 \\
Pa$\beta$~$\lambda$12818  &  12920    &    51. \\
$[$FeII$]$$\lambda$13205 & 13305 &  2.9 \\
Br13~$\lambda$16109    &  16238     &   2.8 \\
$[$FeII$]$$\lambda$16435$^e$   &  16570     &   5.9 \\
Br11~$\lambda$16806    &  16944     &   5.6 \\
Br10~$\lambda$17362    &  17534     &   7.3 \\
Pa$\alpha$~$\lambda$18751   &  18895     &   88.$^d$ \\
H2(1-0)S(3)~$\lambda$19576& 19725    &    5.$^{c,d}$ \\
$[$SIV$]$$\lambda$19635    &  19793     &    4.$^{c,d}$ \\
H2(1-0)S(2)~$\lambda$20332 & 20499   &      2.4$^d$ \\
HeI$\lambda$20581      &  20743     &    7. \\
H2(1-0)S(1)~$\lambda$21218 & 21388   &    2.5 \\
Br$\gamma$~$\lambda$21655    &  21837     &    11.6 \\
\hline \hline
\end{tabular}
\\
$^a$ Observed wavelength in \AA\ .\\
$^b$ Integrated flux in units of $\rm 10^{-15}erg~cm^{-2}~s^{-1}$. For
lines with broad components this column gives the total flux (narrow+broad).\\
$^c$ Flux poorly determined (30--50\% uncertainty)
due to blending with other lines.\\
$^d$ Flux poorly determined (30--50\% uncertainty)
because the line is in a region of bad atmospheric transmission.\\
$^e$ We assume that the narrow line observed at 1.65 $\mu m$ is dominated
by the $[$FeII$]$ line, with little contribution from Br12 (that is much
broader).\\
$^f$ We could not derive a flux for this line due to its sever blending
with H$\alpha$ wings and the [SII] doublet.

\end{table}

\end{document}